\documentclass[review,number,sort&compress]{elsarticle}
\usepackage{lineno}
\usepackage{graphicx}
\usepackage{graphics}
\usepackage{hyperref}
\usepackage{amssymb}
\usepackage{amsmath}
\usepackage{dcolumn}
\usepackage{multirow}

\journal{Nuclear Instruments and Methods A}
\begin{document}
\begin{frontmatter}

\author{Ivan V. Bazarov}
\author{Allen Kim}
\author{Manu N. Lakshmanan}
\author{Jared M. Maxson}

\address{Cornell Laboratory for Accelerator-based Sciences and Education (CLASSE), Cornell University, Ithaca, NY 14853, USA}

\title{Comparison of DC and SRF Photoemission Guns For High Brightness High Average Current Beam Production}
\begin{abstract}
A comparison of the two most prominent electron sources of high average current high brightness electron beams, DC and superconducting RF photoemission guns, is carried out using a large-scale multivariate genetic optimizer interfaced with space charge simulation codes. The gun geometry for each case is varied concurrently with laser pulse shape and parameters of the downstream beamline elements of the photoinjector to obtain minimum emittance as a function of bunch charge. Realistic constraints are imposed on maximum field values for the two gun types. The SRF and DC gun emittances and beam envelopes are compared for various values of photocathode thermal emittance. The performance of the two systems is found to be largely comparable provided low intrinsic emittance photocathodes can be employed.

\end{abstract}


\end{frontmatter}

\section{Introduction}
To realize the fullest potential in a range of applications, Energy Recovery Linacs (ERLs) require high brightness electron beams that are currently beyond the state of the art. In addition to very low beam emittances ($\leq1\,\mu$m rms normalized), these sources need to provide high average currents ($\sim100$\,mA). Photoemission guns, whether utilizing a DC high voltage gap or an RF resonant cavity, have become the technology of choice and remain a key component in photoinjectors. Several efforts are underway in the accelerator community to advance the electron source technology towards generating higher average current and lower emittance beams.

Normal conducting RF guns have performed well in pulsed applications, e.g. see \cite{diag_Akre2008}. CW operation tends to be limited to a lower frequency range (below a GHz) \cite{Rimmer2005} and the problems of ohmic wall losses appear prohibitive for the L band frequency range. DC and superconducting RF (SRF) guns are free of this limitation, which allows the excellent vacuum necessary for high quantum efficiency photocathodes.

Both technologies are actively pursued at the moment at a number of laboratories; for an overview refer to \cite{Nishimori2009, PhysRevSTAB.14.024801}. It is important to understand the main limitations in both cases. DC guns are mainly limited by field emission, whereas SRF guns should allow operation with much higher fields. However, the introduction of a photocathode transport system with load-lock into a clean SRF gun environment without unwanted field emission remains a challenge. The implications of beam dynamics are very different for the two gun types as well. Higher accelerating gradient is of advantage in SRF guns for space charge dominated beams. DC guns, on the other hand, are free of time-dependant forces, which allows for small abberations, as well as longer bunches to reduce the effect of space charge forces.

In this paper we present a comparison of the two gun types for the production of low emittance high average current beams from the point of beam dynamics and emittance performance. In simulations, each gun is followed by a short 1.3-GHz accelerating section (existing Cornell ERL injector cryomodule) that takes the beam energy to 10-12\,MeV where the effect of space charge forces on beam emittance are considerably reduced. We use a genetic multiobjective algorithm \cite{Bazarov05_01}, which proved to be a powerful tool in the accelerator design. Additionally, we implemented flexible (adjustable) gun geometries for both DC and SRF guns to allow for lowest emittance production. In each of the two gun types, constraints are imposed in order to obtain a realistic assessment of their performance and its implications on beam brightness. Additionally, we investigate the effects of the intrinsic photocathode emittance, the laser shape, and various emittance diluting mechanisms present in the system.

While both technologies will continue to be developed, this study presents a self-consistent comparison from the beam performance point of view. It is shown that either technology is capable of generating ultra-low emittance beams necessary for the next generation high current and brightness accelerators. The results indicate that successfully implemented SRF guns should allow superior performance for photocathodes with high intrinsic emittance, whereas the two technologies are largely equivalent in emittance when very low thermal emittance photocathodes are utilized \cite{Karkare11_01}.

In what follows, we introduce our numerical method and explain the variable geometry of the guns as well as the photoinjector beamline used to compare the two technologies. Following the presentation of the main results, we investigate various emittance limiting and degrading mechanisms in both DC and SRF gun based photoinjectors.

\section{Numerical Method}
For the purpose of this study we explore average currents delivered out of each gun of up to 200\,mA, or 154\,pC/bunch at 1.3\,GHz beam repetition rate, with pulses of 0.9\,mm rms bunch length (3\,ps) at the end of the photoinjector for either gun choice. Beam dynamics in photoinjectors at such charge and bunch duration is dominated by space charge phenomena, and experimentally benchmarked codes are essential to understand beam performance implications. There has been an effort in the accelerator community to benchmark the space charge codes and overall good agreement between beam measurements and simulations exist (for example, see \cite{Bazarov08-01}). We implemented genetic algorithm optimizer to use two different space charge codes: \textsc{gpt} (3D) \cite{Greer07-01} and \textsc{astra} (2D radially symmetric) \cite{Flottman00-01}, which demonstrate excellent agreement between each other and the experimental measurements. Due to the axial symmetry of all the beam elements in the studied photoinjector and in the interest of efficiency, the results presented in this paper were obtained with \textsc{astra}.

\subsection{Optimizer Structure}
Our previous work \cite{bazarov05-01} introduced a genetic multi-objective optimization for the photoinjector design. The main advantage of this method is that optimal fronts are obtained, which show the tradeoffs and dependencies in various parameters. This is to be contrasted with a single point conventional design approach (e.g.~a single bunch charge). Detailed space charge simulations are computationally expensive and as previously, a computer cluster is used in these studies.

An important addition to the optimizer is its newly implemented ability to vary the fieldmaps of individual accelerator elements. Pre-computed fieldmaps from a parameterized geometry of an element (DC or SRF gun in this case) are combined in such a way as to allow the generation of new fieldmaps corresponding to new shapes. This process is controlled by the optimizer in minimizing the figures of merit.

Our optimization package is a set of codes that modularizes the optimization process.  The optimization process has two main components: the selector and the variator \cite{bltz2003a}. The algorithm begins with the selector forming a trial set of decision variable solutions that the variator then uses in either \textsc{astra} or \textsc{gpt} simulations of the beamline, the results of which are returned to the selector.  Then the selector chooses the ``fittest'' solutions from the set, based on several (typically two) criteria, known as ``objectives".

\begin{figure}[tbph]
\begin{center}
	\includegraphics[width=1.0\columnwidth]{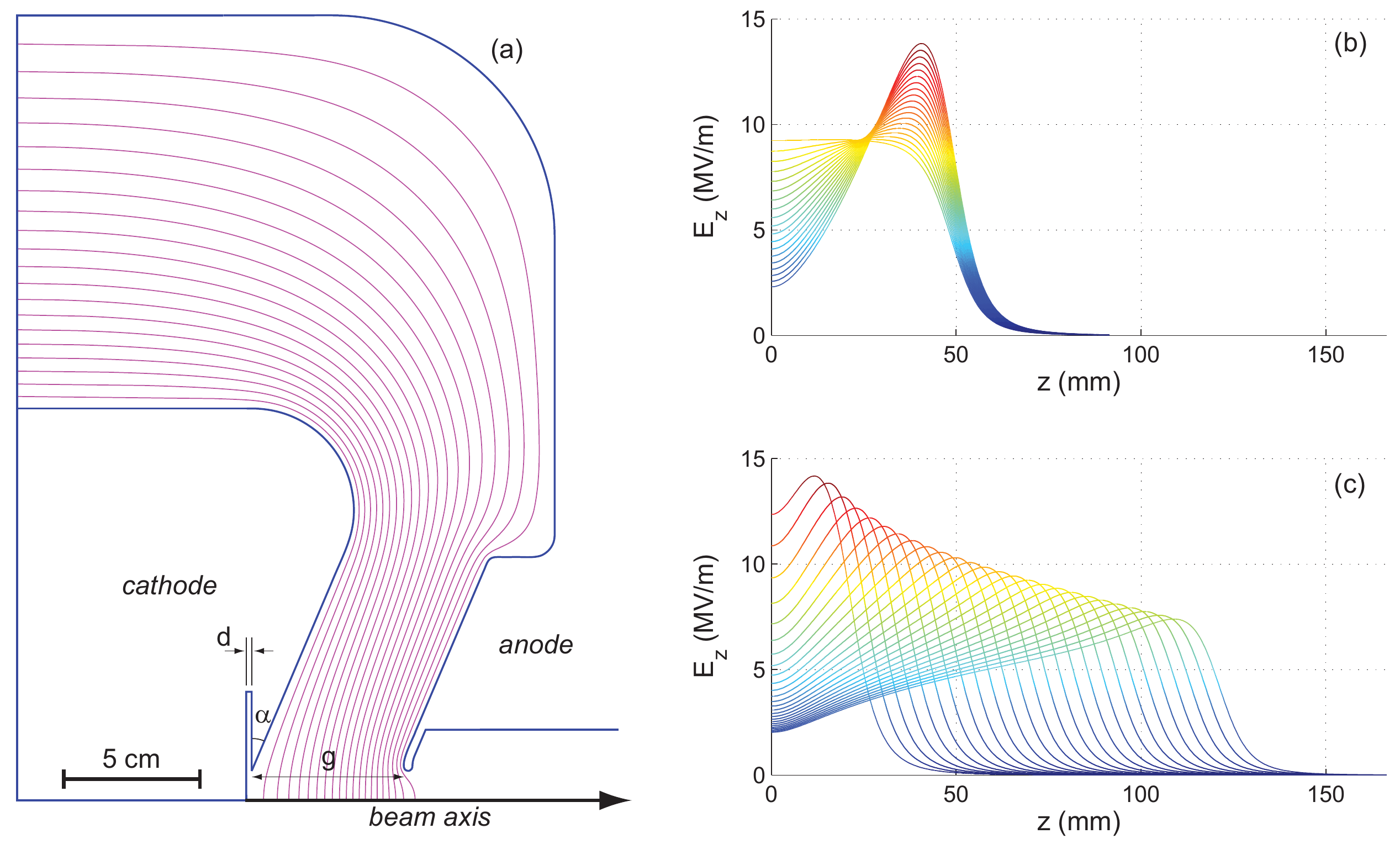}
\caption{(a) The parameterized DC gun geometry: the electrode angle $\alpha$, the cathode-anode gap $g$, and the photocathode recess $d$ are the parameters varied by the optimizer. Equipotential lines are shown. (b) The effect of varying the angle $\alpha$ (from 0 to 45$^\circ$) on the axial electric field for a fixed gap $g=48$\,mm. (c) The effect of varying the gap $g$ (from 20 to 120\,mm) for a fixed $\alpha=23^\circ$.\label{fig:dcgeom}
}
\end{center}
\end{figure}

To form a new trial set for the next generation of solutions, the selector applies two operators to the selected fittest solutions of the previous generation:  (1) ``crossing'', or ``mating'', of two or more solutions; and (2) slightly perturbing (``mutating'') each solution to form a new solution (``offspring'').  The process is then repeated with the newly formed set of offspring solutions and continues for a number of generations, effectively exploring the decision variable space for the best solutions.  In the process, the selector also subjects the solutions to a set of constraints to ensure physically realistic scenarios.  Finally, the algorithm presents a set of optimal solutions as the optimal front. In our study, the objectives are minimum beam emittance and maximum bunch charge, constrained so that the current in the injector does not exceed 200mA, the final bunch duration to be less than 3\,ps rms, and that the fields in DC and SRF guns remain below the physical maxima (detailed below). We expect the minimum emittance solutions to be those with low bunch charge, and thus the inclusion of bunch charge as an objective effectively serves to scan the emittance over the entire range of bunch charges.

The optimizer as a whole will seek to evaluate different solutions with various beam parameters, and more challengingly, solutions with different gun geometries. Field maps for a requested gun geometry could be calculated during optimization; however, we have found it more computationally efficient to calculate field maps for a discrete set of gun geometries prior to the optimization run. These field maps, calculated and post-processed with \textsc{poisson-superfish} \cite{poisson}, are tabulated based on a number of geometry parameters and figures of merit (angles, electric field at the photocathode, peak fields, etc.). The optimizer selects from a continuous space of these geometry parameters, wherein the requested map and its figures of merit are interpolated on the multi-dimensional geometry parameter space.

A powerful addition to the optimizer has been the inclusion of constraints that are any algebraic relationship of the above geometry parameters, figures of merit, or simulation outputs. For instance, this has enabled the implementation of the empirical voltage breakdown condition, which is a power law function relating the DC voltage, and DC gun cathode-anode separation. Furthermore, this capability allows mid-optimization calculation of various functions that depend on both geometry and field map figures of merit.

\begin{figure}[tbph]
\begin{center}
	\includegraphics[width=1.0\columnwidth]{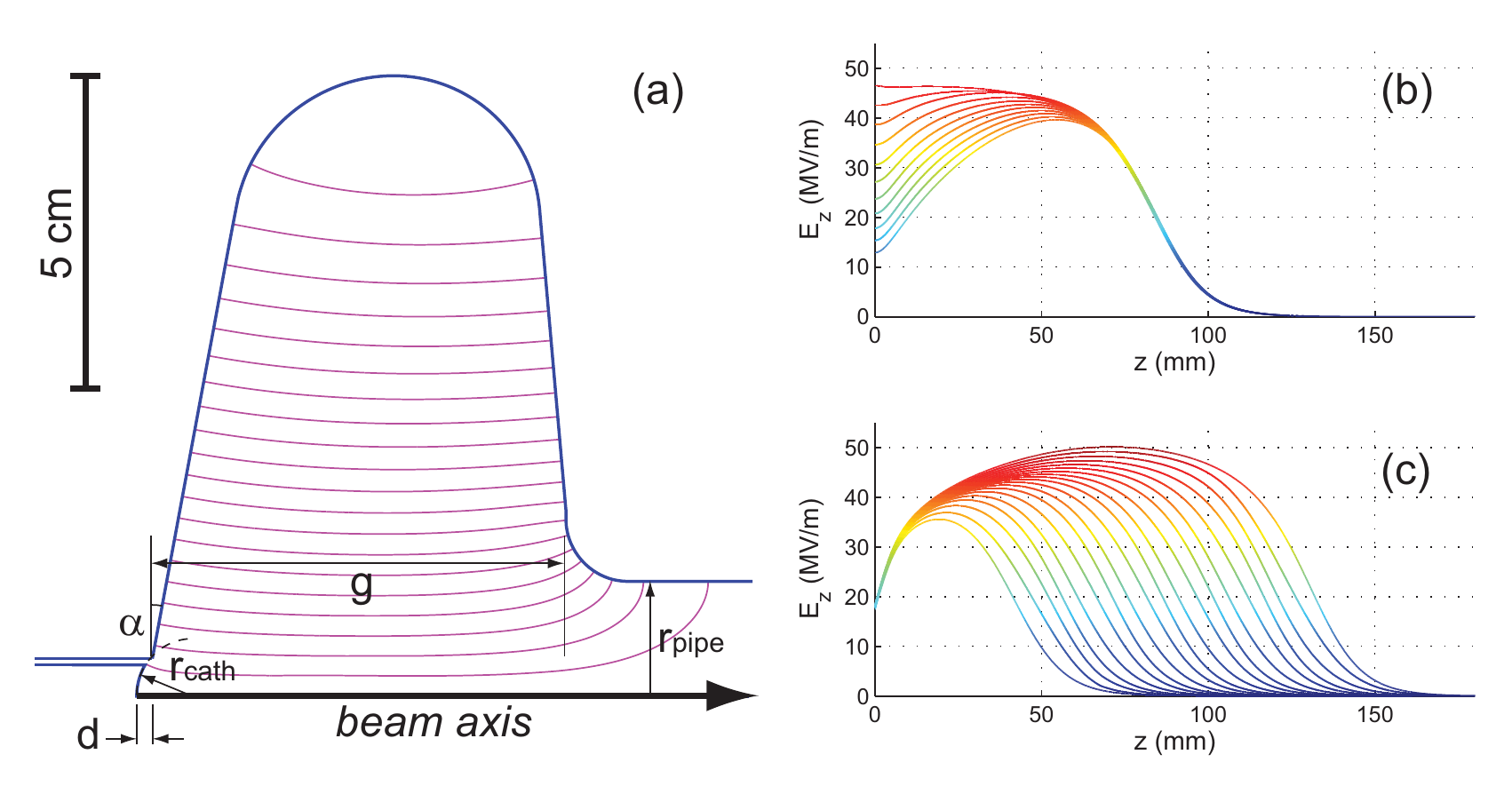}
	\caption{(a) The parameters of the SRF cavity geometry, modeled after the 1.3\,GHz TESLA design. The radius $r_\mathrm{cath}$ and recess $d$ of the photocathode, the angle $\alpha$ of the leftmost cavity wall, the gap $g$ between photocathode and exit pipe, and the exit pipe radius $r_\mathrm{pipe}$ are varied,  with the equatorial cavity radius left as the free parameter for frequency tuning. Here, lines of constant azimuthal magnetic field are shown. (b) The effect of varying the angle $\alpha$ (from 0 to 25$^\circ$) on the axial electric field for fixed $g=75$\,mm, $r_\mathrm{pipe}=1.83\,$cm, $d=0$ and $1/r_\mathrm{cath}=0$. (c) The effect of varying the gap $g$ (from 15 to 105\,mm) for fixed $\alpha=10.5^\circ$, $d=1.67\,$mm, $r_\mathrm{cath}=15\,$mm, and $r_\mathrm{pipe}=1.83\,$cm.\label{fig:srfgeom}}
\end{center}
\end{figure}

\subsection{DC Gun Geometry Parameters}
The DC gun geometry parameters that are varied are the Pierce electrode angle, the cathode anode gap, and the the photocathode recess, as shown in Fig.~\ref{fig:dcgeom}. The cathode recess has an effect of fine-tuning electrostatic focusing at the photocathode. At the end, however, recess was found to be a relatively unimportant parameter for the final injector performance. The gun voltage is also varied directly, being only limited by vacuum breakdown (Figs.~\ref{fig:dcgeom}b and \ref{fig:dcgeom}c depict the highest allowable voltages).

In our optimizations, the gap was allowed to vary from 2-12\,cm, the angle between 0-45$^\circ$, and the recessed between 0-2\,mm. There are a number of emittance tradeoffs when varying gun geometry. An increased angle provides greater focusing, beneficial to counteract space charge, but also decreases the field at the photocathode surface. Decreasing the gap will strengthen the field at the photocathode surface, but will also increase the intrinsic effect of anode defocusing. The voltage and gap will be ultimately limited by the vacuum breakdown limit, to be discussed in Section \ref{sec:breakdown}.

%
\begin{figure}[tbph]
\begin{center}
	\includegraphics[width=0.9\columnwidth]{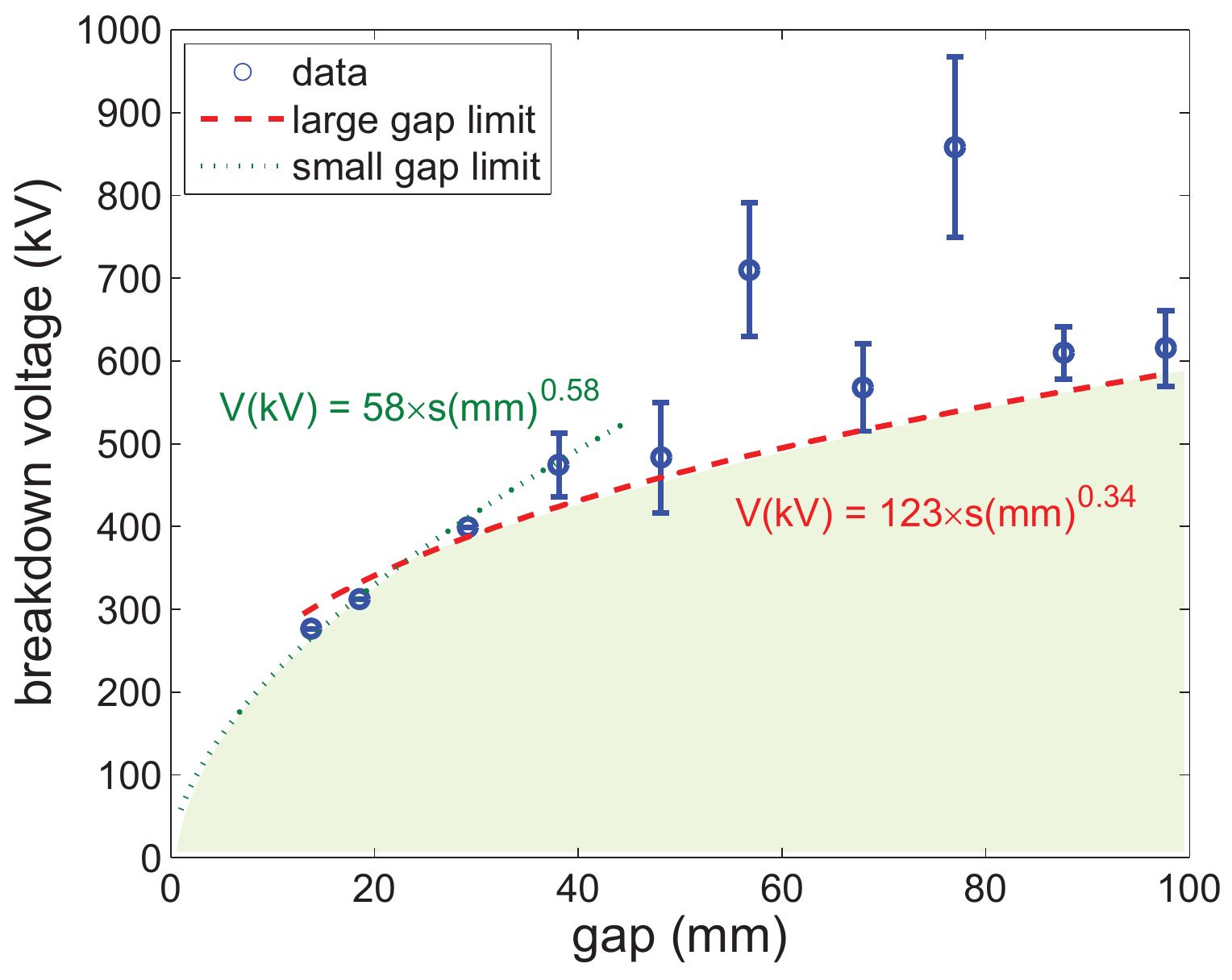}
	\caption{Empirical data of voltage breakdown as a function of gap for planar electrodes, after \cite{Slade08_01}, including fits used in constraining solutions. Filled area shows allowed values. Here, $s$ is defined as the shortest distance between the cathode and anode.\label{fig:slade}}
\end{center}
\end{figure}

\subsection{SRF Gun Geometry Parameters}
We use a one- (or half-) cell SRF cavity design. While it might be beneficial to use multiple cells, our choice was motivated by both simplicity and input coupler considerations. The beam energy even after a one-cell SRF gun can approach 2\,MeV in our optimizations, requiring 400\,kW power coupled into 200\,mA beam (the highest average current considered in this study). These power levels become problematic for input couplers at 1.3\,GHz and a larger number of cells has been ruled out. For the very same reason, the energy boosting cavities in the Cornell ERL injector cryomodule design have only 2-cell cavities at a more modest gradient, each equipped with twin input couplers capable of delivering $\sim120\,$kW RF power into the beam.

The 1.3\,GHz SRF gun cavity geometry is shown in Fig.~\ref{fig:srfgeom}. The SRF gun has 5 varied parameters: the photocathode radius $r_\mathrm{cath}$, the angle of the leftmost cavity wall $\alpha$, and to a lesser extent the photocathode recess $d$, will affect the initial focusing, whereas the gap $g$ and exit pipe radius $r_\mathrm{pipe}$ will determine the extent of the cavity field into the exit pipe. The equatorial cavity radius is used as the parameter for frequency tuning, which is iteratively performed for each set of geometry parameters. We have allowed the following parameter ranges: $\alpha$ is varied between 0-50$^\circ$, $g$ between 1.5-13.5\,cm, $d$ from 0-2.5\,mm, $1/r_\mathrm{cath}$ between 0-0.1\,mm$^{-1}$, and $r_\mathrm{pipe}$ from 0.8-3.9\,cm. Not all geometries within the scanned space can be tuned to 1.3\,GHz or are even possible; however, the successfully generated geometries were seen to form a connected set, as expected. In Fig.~\ref{fig:srfgeom}b and ~\ref{fig:srfgeom}c, the maximum surface electric and magnetic fields are constrained to be equal to the values found in TESLA 9-cell cavity structure \cite{PhysRevSTAB.3.092001} at $E_\mathrm{acc}=$25\,MV/m.

\subsection{Vacuum Breakdown and Critical Fields}
\label{sec:breakdown}
It is essential to constrain maximum fields achievable in respective gun types for a meaningful comparison. For the case of the DC gun, a fundamental limitation is vacuum breakdown precipitated by field emission. In addition to the material choice, surface preparation as well as the area and the gap separating the electrodes play an important role. While ceramic puncture is the present limitation in raising gun voltage higher for many existing DC guns, technological solutions such as the use of segmented, shielded ceramic \cite{nagai:033304} may entirely mitigate the puncture problem. In this case, the emphasis is shifted towards the fields in the beam region of the cathode-anode gap. While the field emission current scaling is well known via the Fowler-Nordheim relations, field emission sites, often caused by inclusions within the electrode material, are highly stochastic in concentration, and cross-talk mechanisms between the anode and the cathode (e.g.~x-ray generation, electron-induced gas desorption, etc.) make the onset of the field emission notoriously difficult to predict. However, empirical data have been collected in \cite{Slade08_01} concerning vacuum breakdown voltage as a function of gap, which is plotted for our region of interest in Fig.~\ref{fig:slade}. In the figure, $s$ is the shortest distance between the cathode and the anode, approximately given by $s\approx g \cos(\alpha)$. The breakdown voltage is computed for each combination of the gun geometry parameters and if the gun voltage exceeds the breakdown voltage, that trial solution is invalidated.

SRF guns are also prone to field emission problems \cite{PhysRevSTAB.14.024801}. One important challenge is an introduction of the photocathode (via a load-lock) into the ultra-clean SRF cavity environment. A number of SRF guns have displayed high levels of the field emission, which is especially significant when high quantum efficiency materials are present in the system. We use an optimistic criterion with fields being limited by the standard TESLA cavity parameters at $E_\mathrm{acc} = 25$\,MV/m, which will undoubtedly be more difficult to achieve in an SRF photoemission gun. Both peak electric field $E_\mathrm{pk}$ and magnetic field $H_\mathrm{pk}$ at the niobium surface is calculated for each gun geometry. The following requirements are imposed during simulations $E_\mathrm{pk}/E_\mathrm{acc}\leq 2$ and $H_{pk}/E_{acc}\leq4.26$\,mT/(MV/m) \cite{PhysRevSTAB.3.092001}.  We find that the majority of solutions within our parameter space were limited by the restriction on the surface electric field ($<50$\,MV/m).

\begin{figure}[t!]
	\includegraphics[width=\columnwidth]{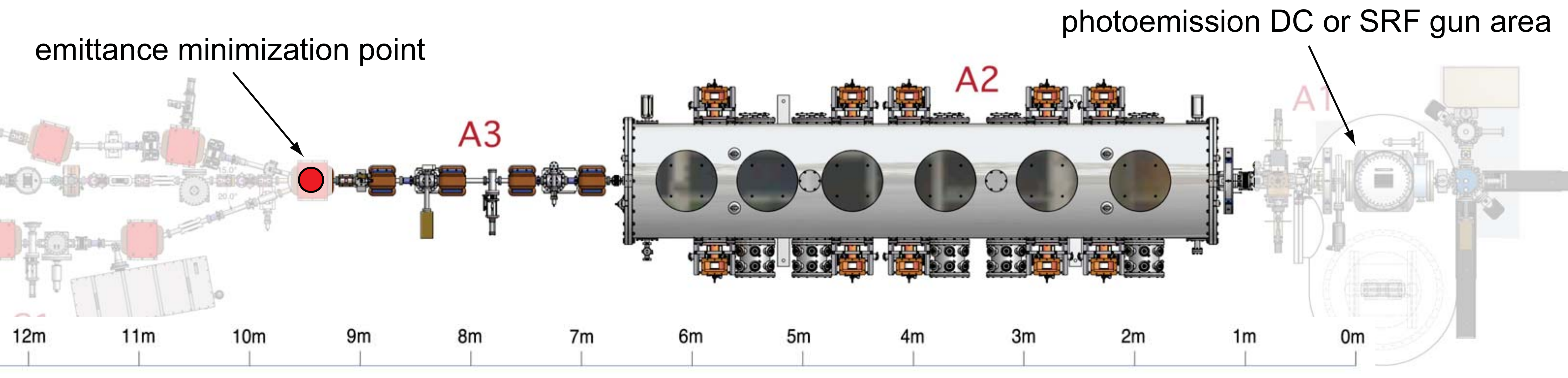}
	\caption{Injector layout modeled after Cornell ERL electron source. Beam direction is to the left.\label{fig:l0inj}}
\end{figure}

\section{Optimization Parameters}

\subsection{Beamline}
Both the SRF and DC beamlines are modeled after the existing Cornell ERL injector, a schematic of which is shown in Fig.~\ref{fig:l0inj}. The DC beamline has no modifications to the Cornell injector, which includes two emittance compensating solenoids with a normal conducting RF bunching cavity, followed by a 5-m long cryomodule with five 2-cell 1.3 GHz SRF cavities, and then a drift section until the emittance measurement system at $z=9.5\,$m from the photocathode.

\begin{figure*}[bth]
	\includegraphics[width=\textwidth]{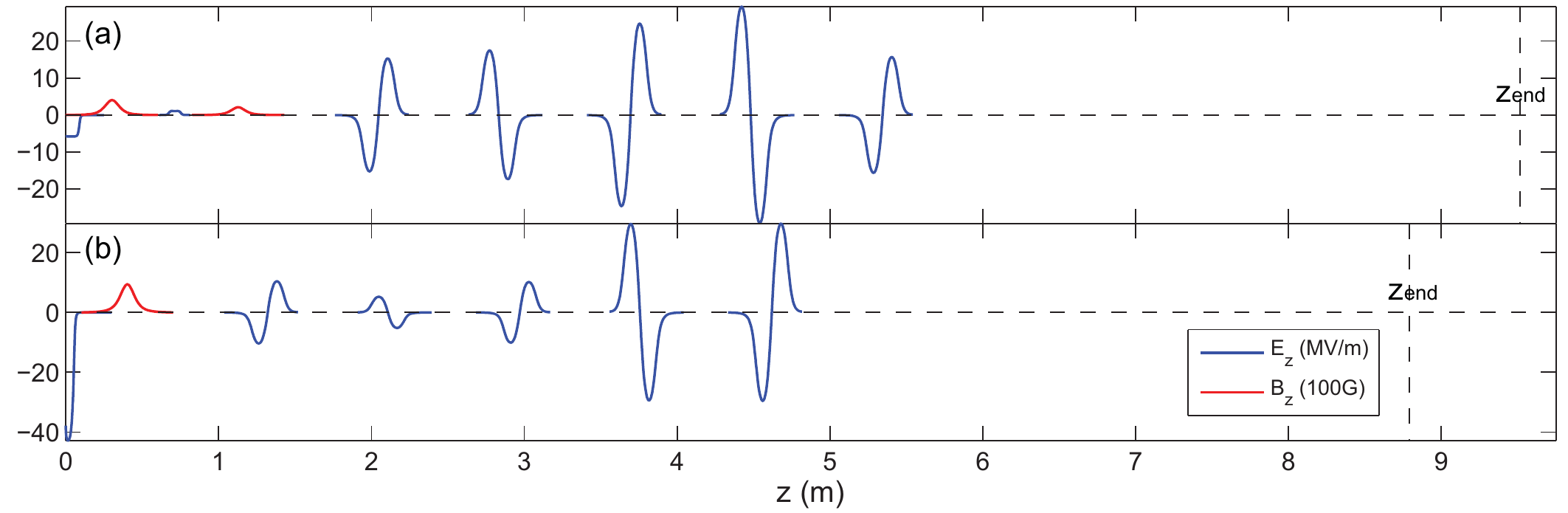}
	\caption{Typical axial fields in the injector: (a) the DC and (b) the SRF gun-based.\label{fig:fieldmaps}}
\end{figure*}

The buncher cavity is operated at zero-crossing and is used to compress the electron bunches. This works very well due to the low energy out of the DC gun requiring only modest buncher fields. However, in the case of the SRF gun, owing to the higher gun energy, the buncher is of limited utility. Thus, it was completely eliminated, and the beamline for the SRF gun based photoinjector has only one solenoid between the SRF gun and the energy boosting cryomodule. We have chosen a distance of 40\,cm between the solenoid center and the gun photocathode to allow for sufficient magnetic field attenuation at the niobium structure.

Refer to Fig.~\ref{fig:fieldmaps} for an example of axial fields for both photoinjector types. Each magnet current, cavity phase, and amplitude are varied by the optimizer. All the beamline elements can adopt a range of values that have been demonstrated in the Cornell ERL photoinjector (e.g.~the maximum electric field on axis in SRF 2-cell cavities stays below 30\,MV/m, while the RF buncher does not exceed $\sim2$\,MV/m).

\begin{figure}[t!]
\begin{center}
	\includegraphics[width=0.8\columnwidth]{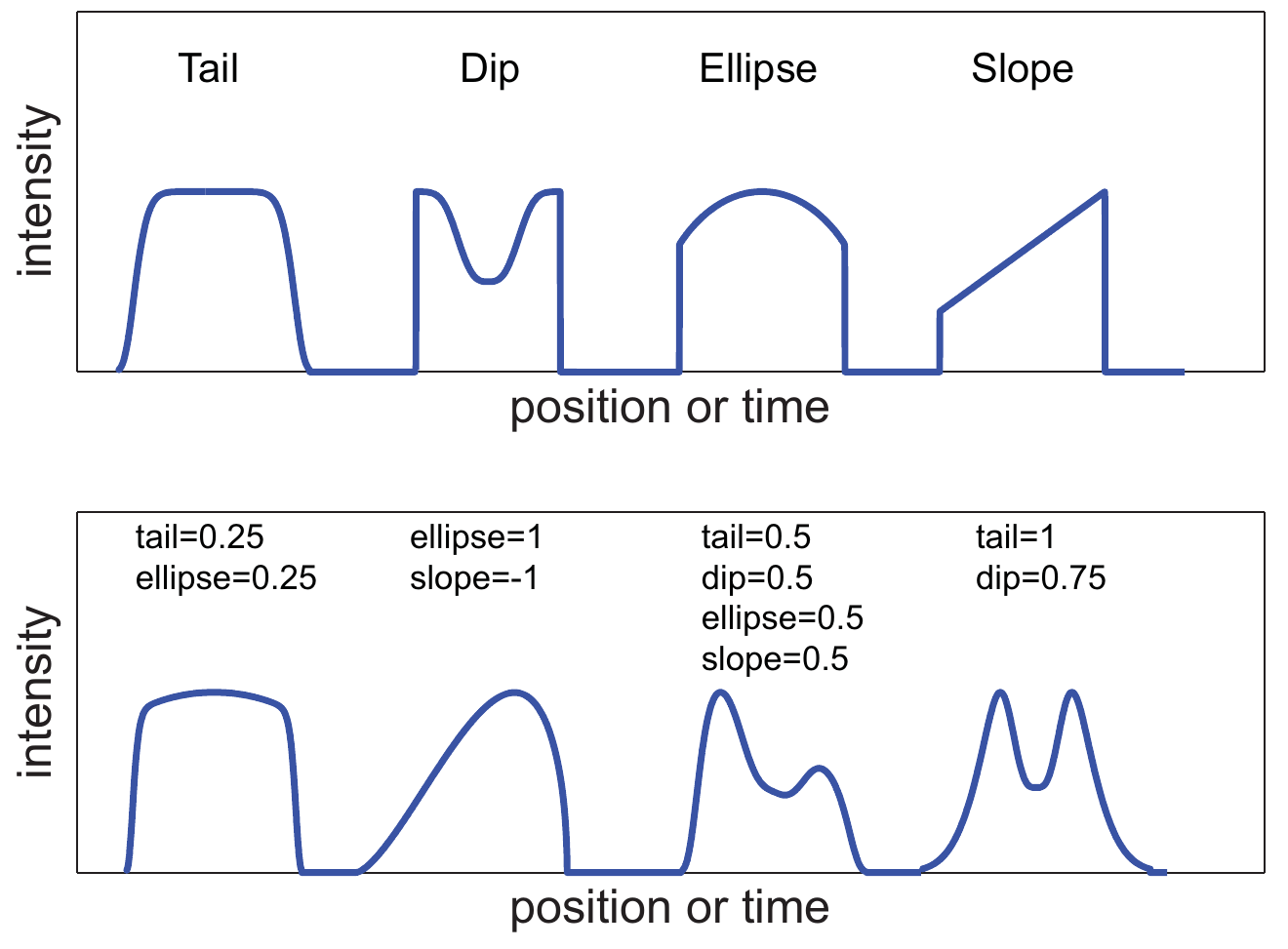}
	\caption{Above: Archetypal laser pulse shapes, variably superimposed to form the requested laser pulse shape. Below: Sample pulse shapes including their superposition parameters.\label{fig:shapes}}
\end{center}
\end{figure}

\subsection{Photocathode and Laser Shaping}
Photocathode properties play an important in production of high brightness electron beams. The Mean Transverse Energy (MTE) associated with photoemitted electrons along with the cathode electric field set a limit to the highest beam brightness available from a photoinjector \cite{Bazarov09-01}. In terms of the laser rms spot size $\sigma_{xy}$, the intrinsic emittance (rms normalized) from the photocathode is given by
\begin{equation}
\epsilon_{nxy,\mathrm{th}}=\sigma_{xy}\sqrt{\frac{\mathrm{MTE}}{mc^2}},
\end{equation}
where $mc^2$ is the electron rest energy. Additionally, photoemission response time impacts the effective use of laser shaping. Several photocathode materials hold immediate promise. $\mathrm{K_2CsSb}$ has good quantum efficiency at a convenient laser wavelength (green) and additionally demonstrates good longevity for high average current applications. Its exact value for MTE is still under investigation. GaAs features very low $\mathrm{MTE}=0.12$\,eV at 520\,nm and a prompt \cite{Bazarov08_02} response ($<1$\,ps). In this study we use 3 values for MTE: 0.5\,eV, 0.12\,eV, and 0.025\,eV \cite{Karkare11_01}.

To achieve very small emittances it is essential to control space charge forces via laser shaping. For a DC beam, a transverse flat-top distribution is ideal as it generates linear space charge forces that do not increase beam emittance. For beams in free space, a uniform density 3D ellipsoid gives a linear force in any direction. The conducting boundary condition at the photocathode surface changes this idealized picture. Additionally, the space charge forces can couple transverse and longitudinal motion. We have included several parameters to optimize the temporal profile of the laser pulse by allowing a wide range of pulse templates to explore effective laser shapes from the electron beam dynamics point of view. These pulse templates are shown in Fig.~\ref{fig:shapes}. We have allowed the laser pulse duration to vary between 0 and 30\,ps. The longer bunch lengths near the gun allow for reduced density of space charge, and thus it is expected that the optimizer will push for long pulses, up to the limits set by RF-focusing induced emittance growth. This is in fact the situation we observe in the DC gun case, whereas the SRF gun case never exceeded laser pulse duration of 10\,ps rms. The final bunch duration in all cases is constrained to be less than 3\,ps rms, primarily driven by the considerations of limiting induced energy spread from long bunches in the main linac of ERL \cite{Bazarov03-01}.

\begin{figure}[b!]
\begin{center}
	\includegraphics[width=\columnwidth]{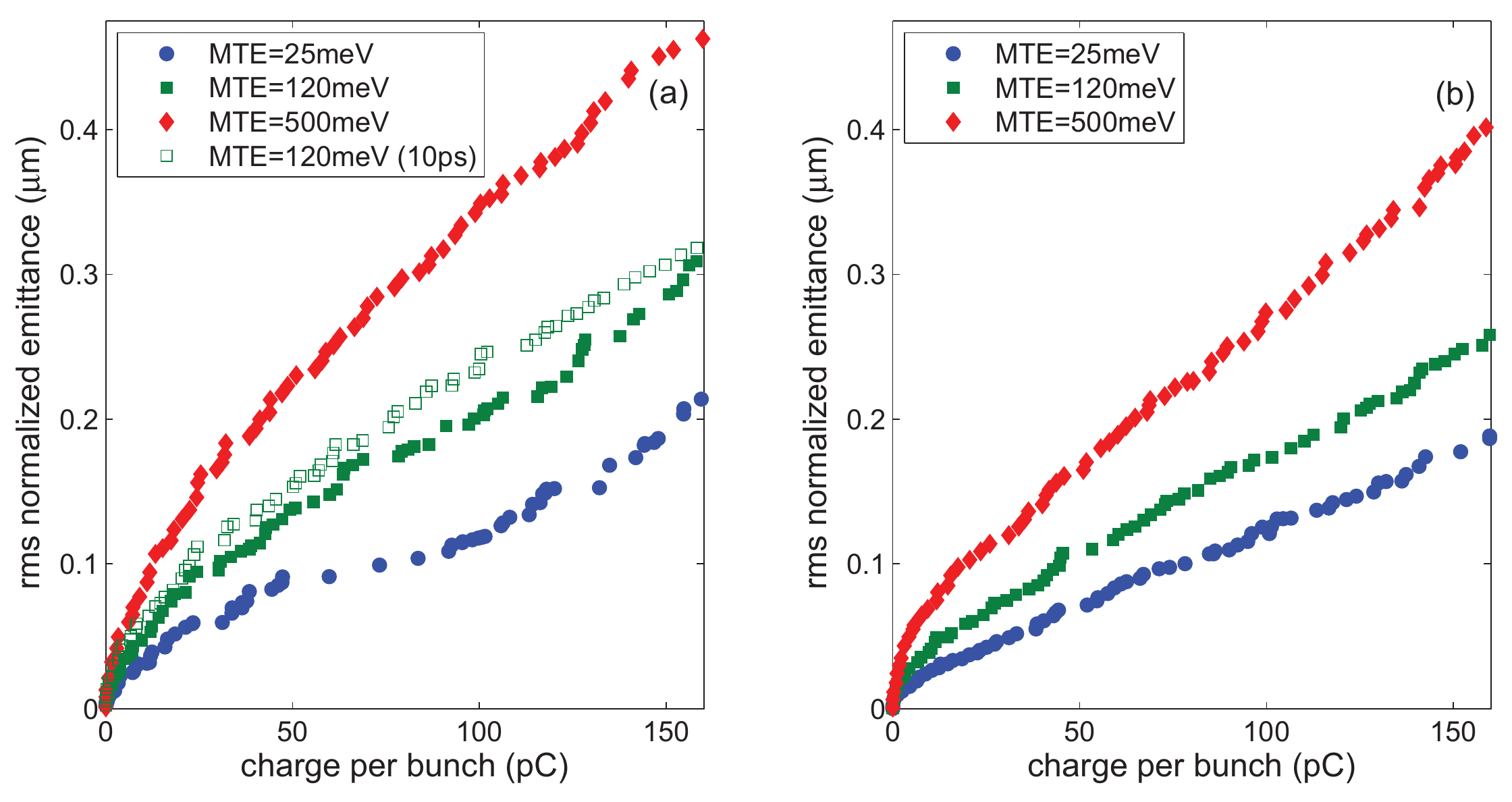}
	\caption{Optimized emittance for various bunch charges for (a) DC gun and (b) SRF gun based photoinjectors.\label{fig:emittance}}
\end{center}
\end{figure}

\section{Results}
\subsection{Final Emittance}
\begin{table}
\caption{\label{tab:optim}Main injector parameters after optimization.}
\begin{tabular}{lcc}
Parameter&DC gun&SRF gun\\
\hline
Charge & $80$\,pC & $80$\,pC\\
Laser spot size (rms) & $0.35$\,mm  & $0.21$\,mm\\
Laser pulse (rms) & $10$\,ps & $9$\,ps\\
Thermal emittance (rms) & $0.17\,\mu$m & $0.10\,\mu$m\\
Cathode field ($t=0$) & 5.1\,MV/m & 16.6\,MV/m\\
KE after the gun& $0.47$\,MeV & $1.91$\,MeV\\
Buncher peak field & $1.2$\,MV/m & -- \\
SRF cavities1,2 peak $E_z$ & $20,22$\,MV/m & $11,6$\,MV/m\\
SRF cavities1,2 phase & $-25,-37^\circ$ & $-60,-40^\circ$\\
Solenoid1 peak field & $0.038$\,T & $0.094$\,T\\
Solenoid2 peak field & $0.023$\,T & -- \\
\hline
Transverse emittance (rms) &   $0.21$\,$\mu$m     & $0.15$\,$\mu$m\\
Bunch length (rms) & $0.89$\,mm & $0.86$\,mm\\
Longitudinal emittance (rms) & $8.2$\,mm-keV & $9.2$\,mm-keV\\
Kinetic energy & $12.4$\,MeV & $10.3$\,MeV\\
\end{tabular}
\end{table}


Results for beam emittance from the DC and SRF gun-based photoinjectors are presented in Fig.~\ref{fig:emittance}. Each injector type shows the results for different photocathode MTE values. We note that the SRF gun performs better for larger MTE values, whereas the results are essentially identical for $\mathrm{MTE}=0.025$\,eV. In what follows, the $\mathrm{MTE}=0.12$\,eV case with 80\,pC/bunch is studied in more detail. The laser duration for the DC gun is pushed to the longer limit (30\,ps rms) in the optimization while the bunch length is being compressed to 3\,ps at the end of the beamline without noticeable emittance degradation. Pulse stacking with birefringent crystals is very effective in generating longer pulses and allows a degree of control of the laser temporal profile \cite{Bazarov08_01}. Generating 30\,ps rms laser pulses with fast rise and fall times may prove challenging. Therefore, the laser pulse duration was constrained to 10\,ps rms in one of the optimizations for the DC gun. The results (Fig.~\ref{fig:emittance}a) show that the final emittance is not very sensitive on the initial laser pulse duration owing to the presence of RF buncher cavity. In what follows, we compare DC and SRF guns for similar initial laser pulse durations (10 and 9\,ps rms respectively).

The main photoinjector parameters for the two gun cases are given in Table~\ref{tab:optim}. The gradients of the first 2 SRF cavities and their phases are critical parameters and are given in the table. It is seen that large off-crest phase values are chosen for gradual bunch compression (more so for the SRF gun case without a dedicated buncher cavity). The subsequent cavities are less critical and their phases can be chosen more freely, e.g.~from considerations of removing correlated energy spread in the bunch.

\subsection{Optimal Geometries}
\begin{figure}[t!]
\begin{center}
	\includegraphics[width=\columnwidth]{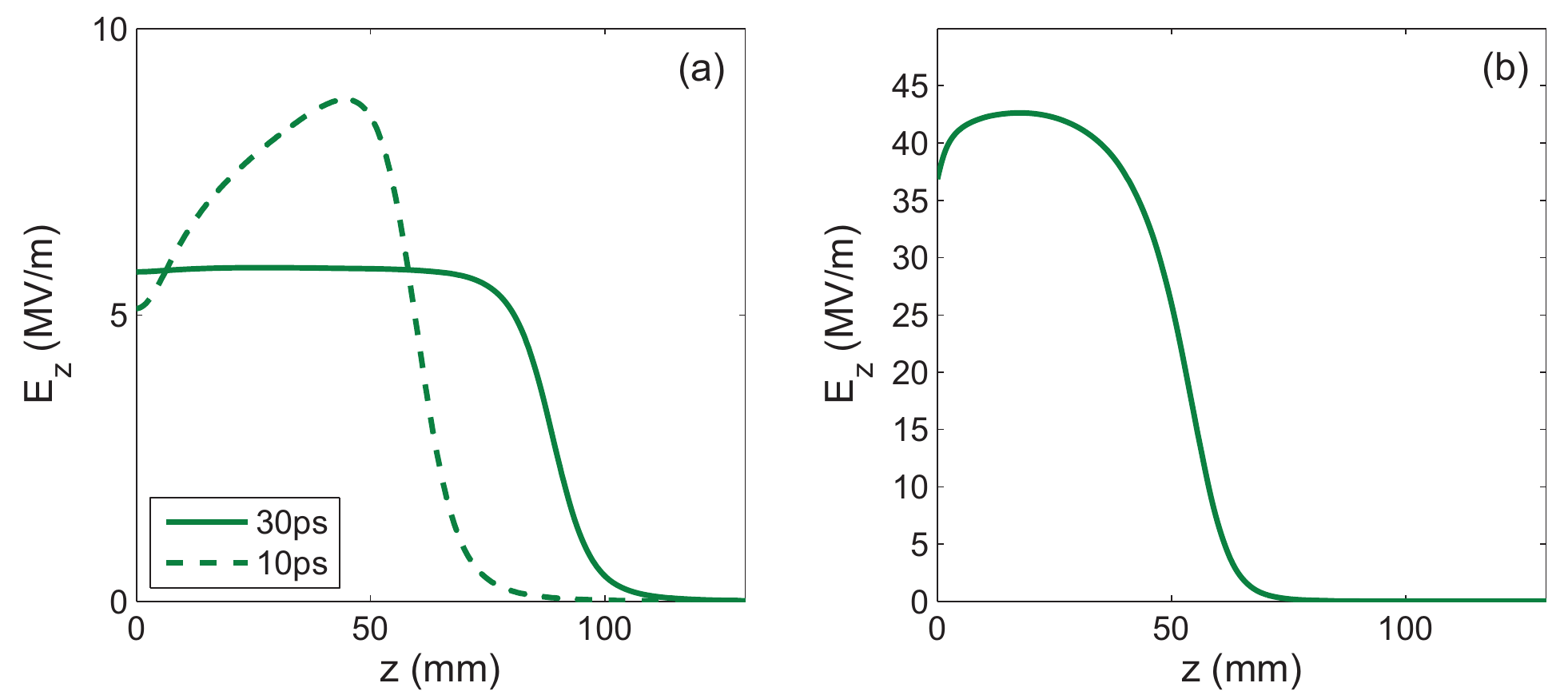}
	\caption{Field profiles for optimized geometries: (a) DC and (b) SRF guns.\label{fig:guns}}
\end{center}
\end{figure}

Fig.~\ref{fig:guns} shows the optimized field profiles inside the DC and SRF guns. It is interesting to note that the long laser pulse case (30\,ps), which has a smaller space charge effect, drives the gun geometry towards a flat cathode electrode ($\alpha \approx 0$) and a gap $g \approx 9$\,cm, thereby increasing the photocathode field and the voltage. On the other hand, the shorter laser pulse calls for an additional electrostatic focusing and has a cathode angle of $\alpha \approx 10^\circ$ and a shorter gap of $g \approx 6$\,cm. The gun voltages are 515 and 475\,kV for 30 and 10\,ps rms pulse durations respectively.

The optimized SRF gun geometry for the case presented in Fig.~\ref{fig:guns}b has $\alpha = 2.3^\circ$, $g = 4.4$\,cm, $r_\mathrm{pipe}=0.9$\,cm, $r_\mathrm{cath}=4$\,cm, and no cathode recess. We note that the exit pipe diameter, while always minimized by the optimizer, does not represent a critical parameter and can be enlarged without significant effect on beam emittance.

\subsection{Laser Shaping}
\begin{figure}[b!]
\begin{center}
	\includegraphics[width=0.9\columnwidth]{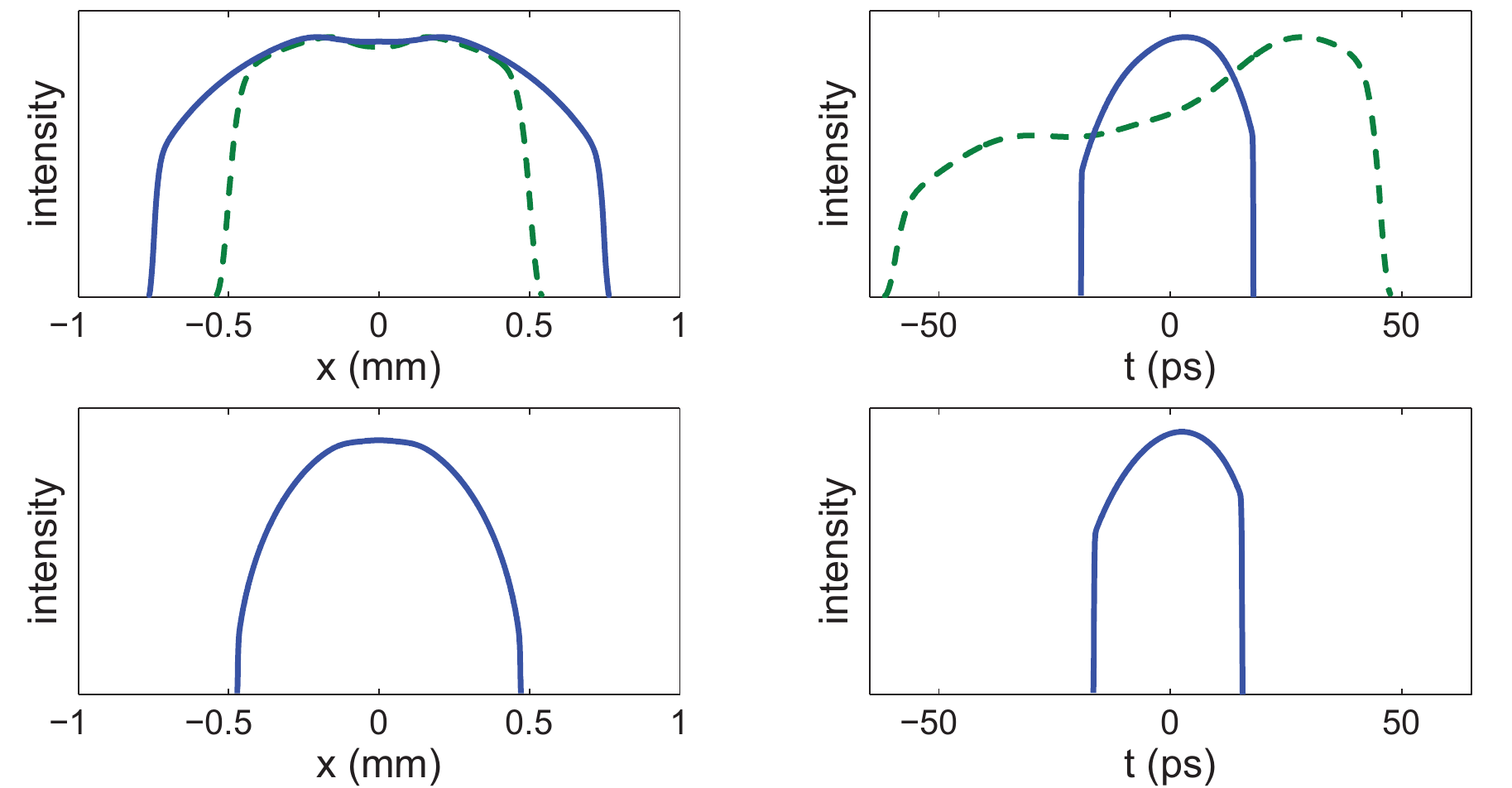}
	\caption{(Top row) The optimal shape for the DC gun for two different laser pulse lengths, 10\,ps (solid) and 30\,ps rms (dashed line). (Bottom row) The optimal shape for the SRF gun.\label{fig:laser}}
\end{center}
\end{figure}

Fig.~\ref{fig:laser} shows the transverse and longitudinal laser profiles selected by the optimizer for 80\,pC bunch operation in the two gun cases. All the profiles are normalized to the same peak value and are shown on the same spatial or temporal scale for comparison. It is interesting to note the asymmetric laser profile in the case of DC gun with the longer pulse, which is used to balance off the asymmetric fields arising at the photocathode near the space charge extraction limit \cite{Bazarov09-01}.

\begin{figure}[t!]
\begin{center}
	\includegraphics[width=\columnwidth]{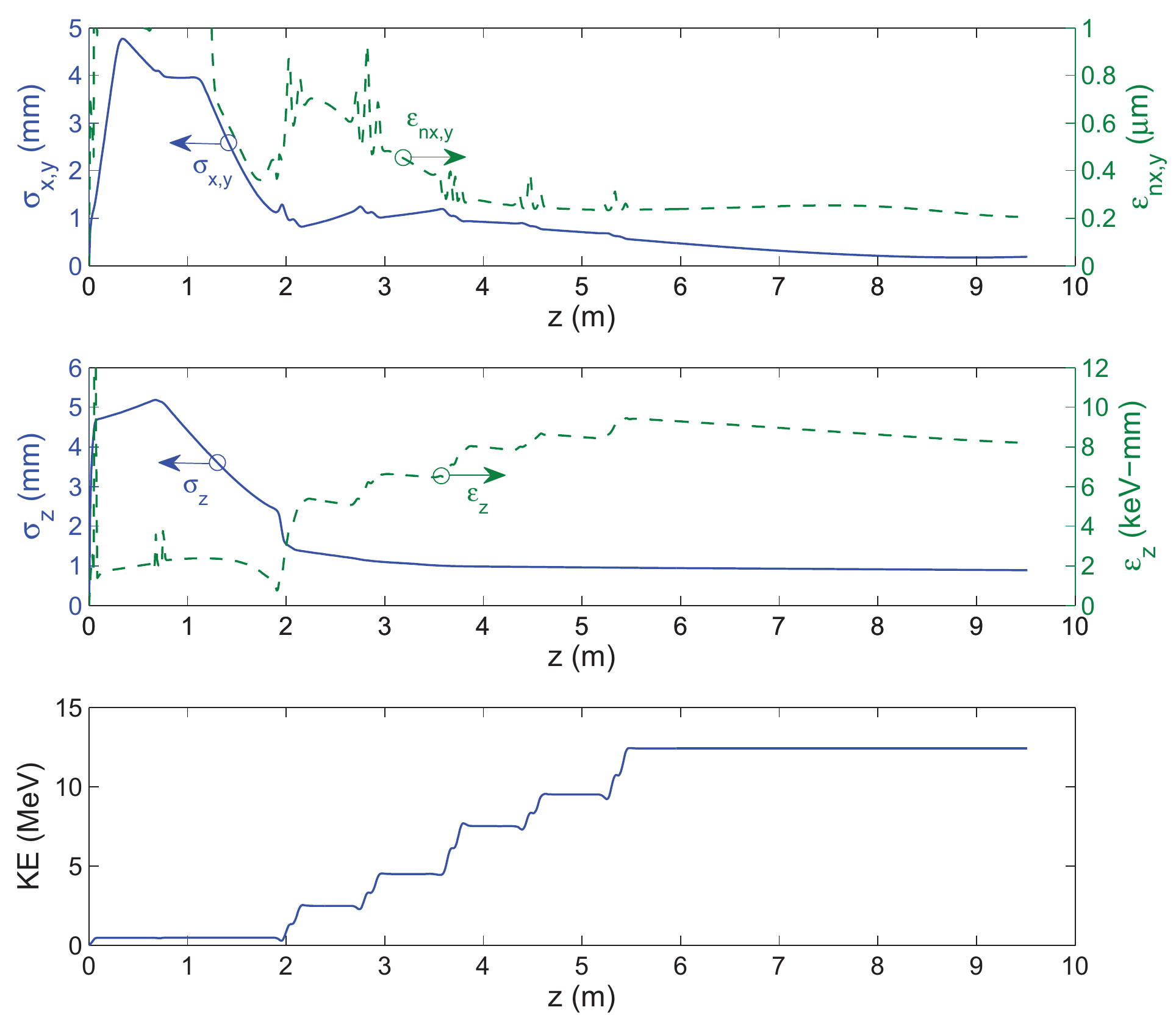}
	\caption{Beam envelopes for 80\,pC ($\mathrm{MTE}=120$\,meV) in the DC gun photoinjector. The initial laser pulse is 10\,ps rms.\label{fig:dcenv}}
\end{center}
\end{figure}

\begin{figure}[b!]
\begin{center}
	\includegraphics[width=\columnwidth]{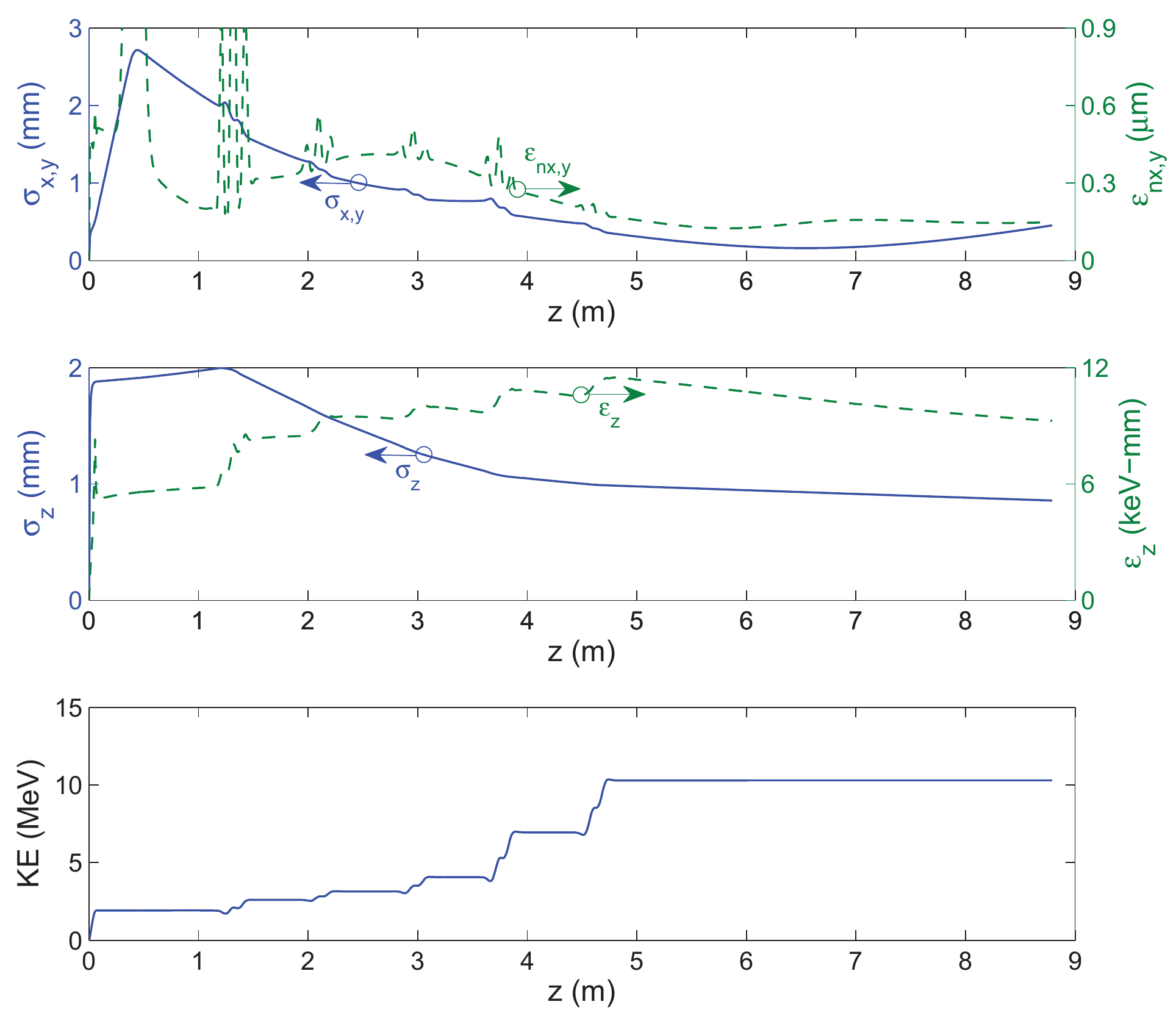}
	\caption{Beam envelopes for 80\,pC ($\mathrm{MTE}=120$\,meV) in the SRF gun photoinjector.\label{fig:srfenv}}
\end{center}
\end{figure}

\subsection{Beam Envelopes}
\begin{figure}[tphp]
\begin{center}
	\includegraphics[width=\columnwidth]{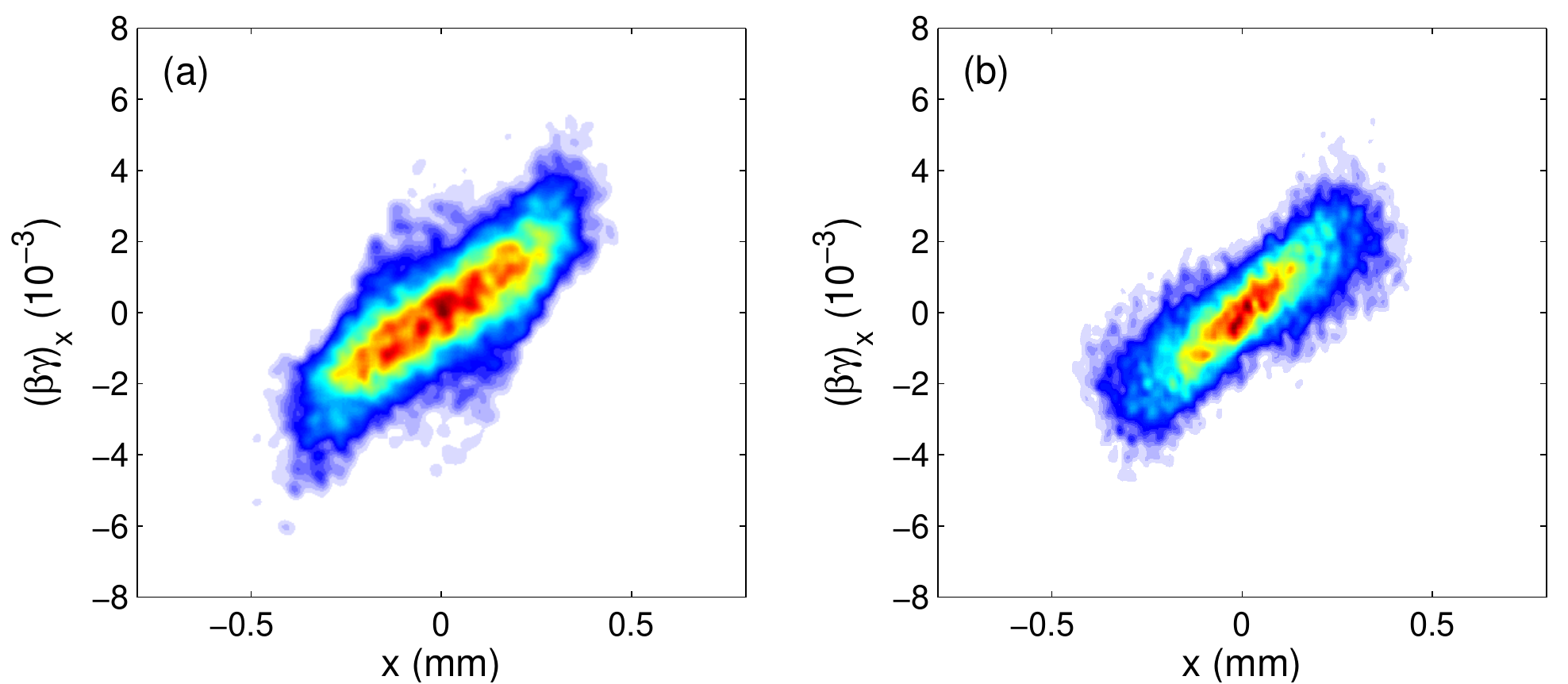}
	\caption{Examples of the final phase space for 80\,pC bunches ($\mathrm{MTE}=0.12\,$eV) for (a) DC and (b) SRF guns.\label{fig:phase}}
\end{center}
\end{figure}

Figs.~\ref{fig:dcenv} and \ref{fig:srfenv} show beam envelopes for the two gun cases with 80\,pC bunches, along with rms transverse and longitudinal emittances, and beam kinetic energy vs.~the longitudinal position. The most salient difference between the two gun types is in $\times2$ larger beam size at the exit of the DC gun as opposed to the SRF gun, as well as a more dramatic bunch length variation along the longitudinal position in the DC gun photoinjector. The final beam parameters, however, end up being quite comparable between the two gun types. Finally, Fig.~\ref{fig:phase} shows the final transverse phase space near the beam waist ($z=9.5$\,m for the DC and $z=7$\,m for the SRF guns).

\subsection{Performance After the Gun}
When commissioning either gun type, it is useful to know the expected gun performance just after the gun exit, since that configuration will be most relevant during initial commissioning. Such a study has been performed for the gun geometries shown in Fig.~\ref{fig:guns}. The shorter commissioning beamline consists of either the DC or the SRF gun with a solenoid placed at 0.3 or 0.4\,m respectively (photocathode to the solenoid center) followed by a 1-m drift to emittance measurement diagnostics.

\begin{figure}[b!]
\begin{center}
	\includegraphics[width=\columnwidth]{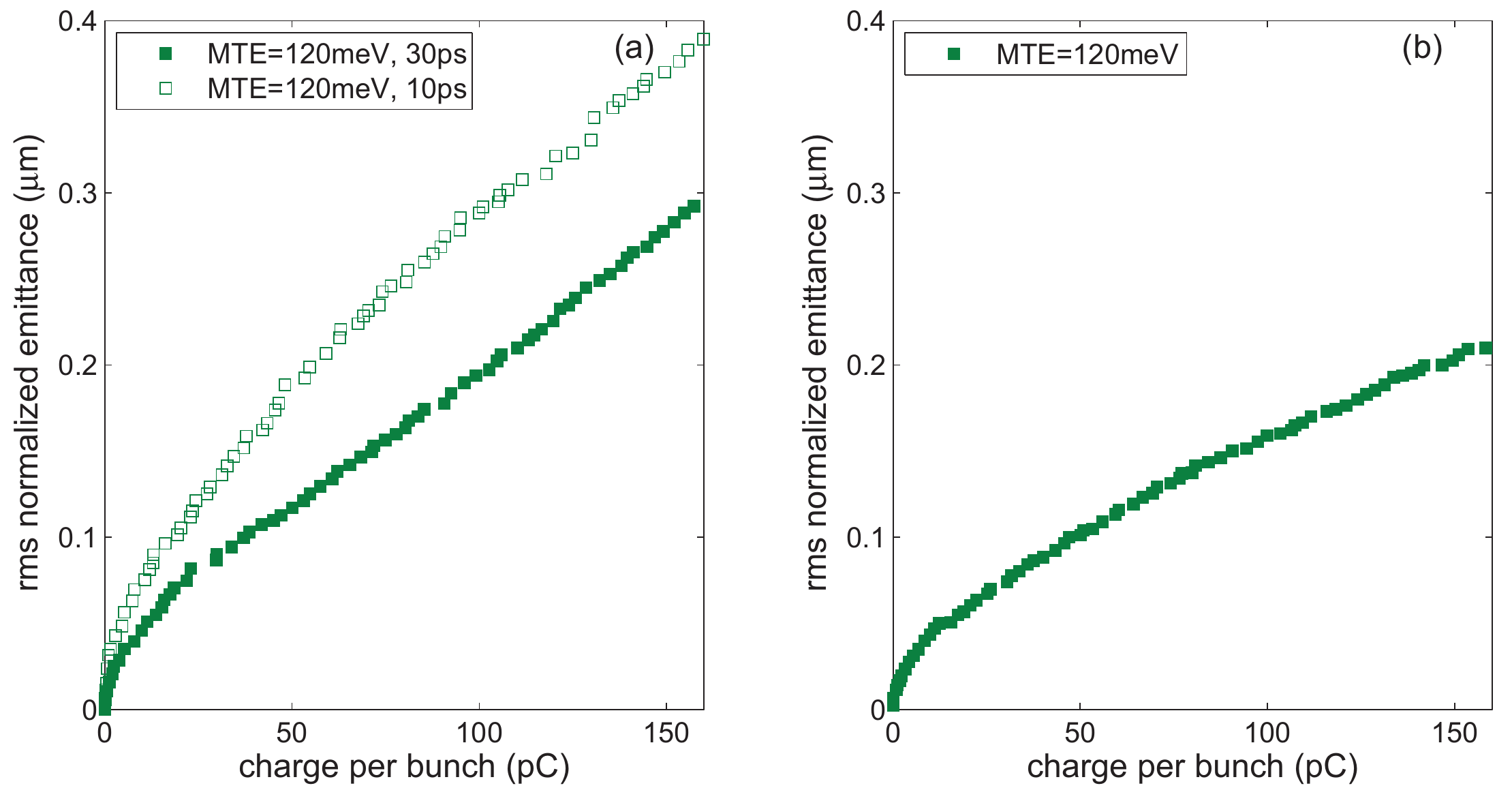}
	\caption{Emittance at the end of the shortened beamline for (a) DC and (b) SRF guns.\label{fig:short}}
\end{center}
\end{figure}

\begin{figure}[tbph]
\begin{center}
	\includegraphics[width=\columnwidth]{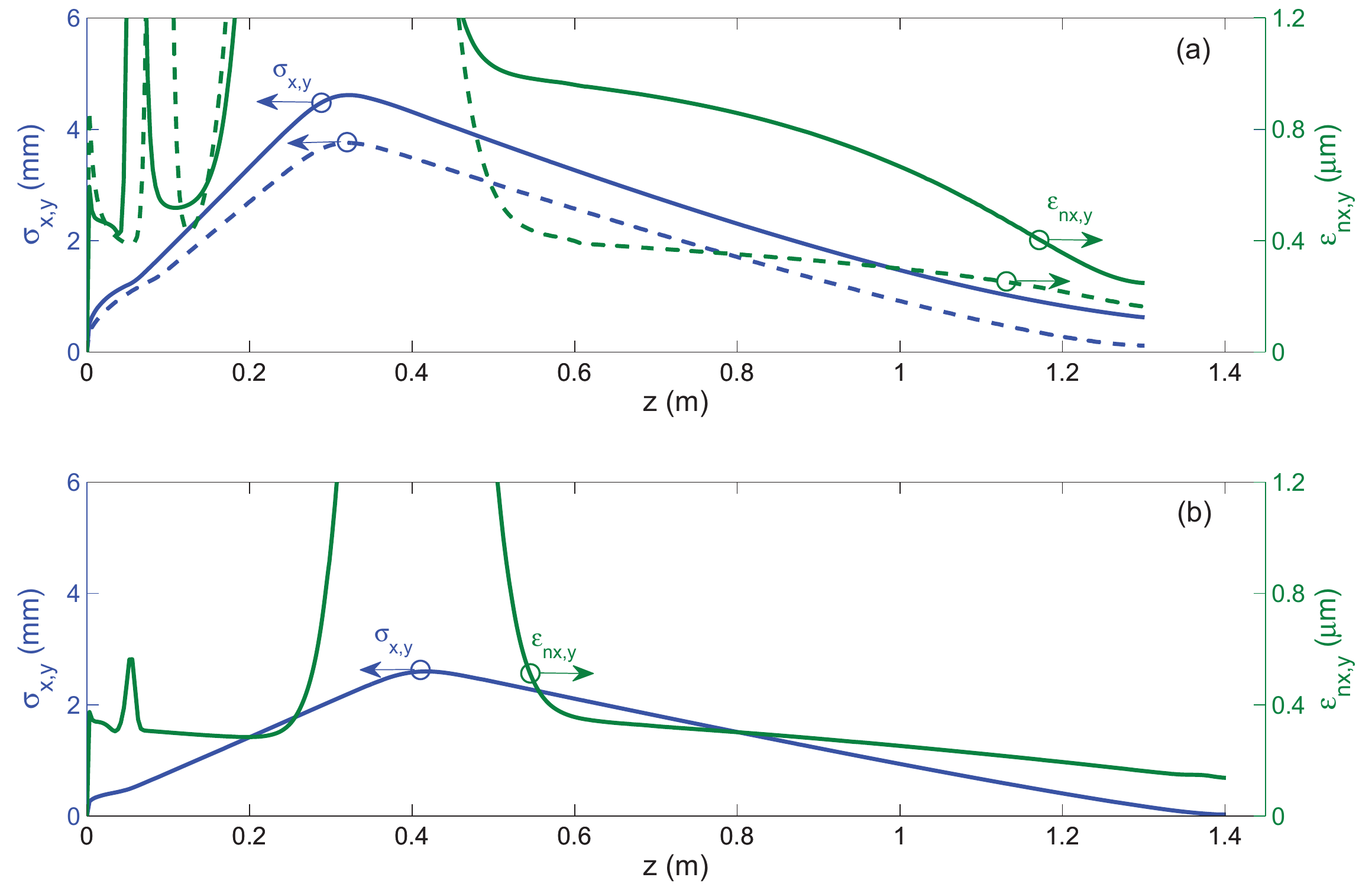}
	\caption{Beam envelopes and emittance in the shortened beamline for (a) DC and (b) SRF guns. The DC gun shows two cases: 10\,ps (solid) and 30\,ps (dashed line) rms initial laser pulse. The photocathode $\mathrm{MTE}$ is 120\,meV.\label{fig:shortenv}}
\end{center}
\end{figure}

The results summarizing emittance performance for such a short beamline are shown in Fig.~\ref{fig:short}. Additionally, Fig.~\ref{fig:shortenv} shows beam envelope and transverse emittance for 80\,pC bunch charge.

It is interesting that the DC gun displays noticeably larger emittance for the 10\,ps rms laser pulse when compared to 30\,ps, as opposed to only a small change seen in the full $\sim10$\,MeV beamline. In this case the larger beam size for the shorter laser pulse causes an increase in the contribution of the solenoid aberrations on final emittance. The question of aberrations and various emittance degrading effects is discussed in detail in the next section.

\section{Discussion}
Achieving a very low beam emittance requires control of many phase space diluting phenomena, including space charge, optics aberrations, and time-dependant (transverse) RF fields. The fact that the CW operation typically requires accelerating fields that are smaller than what can be accomplished in a pulsed accelerator means that the beam dimensions are necessarily larger with correspondingly increased emittance dilution arising from the sampling of field nonlinearities over a larger spatiotemporal volume.

\subsection{Electric Field at the Cathode}
The electric field at the photocathode sets the lower limit on the laser spot size before the onset of the virtual cathode instability \cite{Bazarov09-01}. Together with the photocathode MTE, this decides the smallest achievable emittance.  The relevant figure of merit is the electric field during the electron emission, e.g.~5.1\,MV/m for the DC and 16.6\,MV/m for the SRF guns (see Table~\ref{tab:optim}). The cathode's intrinsic emittance scales as $\epsilon_{nxy,\mathrm{th}} \propto \sqrt{1/E_\mathrm{cath}}$. In case of the very low MTE (0.025\,eV), the final emittance is about 50\% due to the intrinsic photocathode emittance and the other 50\% from residual emittance degrading effects such as the ones discussed below. The fraction of the cathode intrinsic emittance of the final value becomes larger for higher MTE values. We also note that for the smallest MTE value (0.025\,eV), the actual laser spot size difference between the two cases (DC and SRF) is less than what naively might be expected from $E_\mathrm{cath}$ scaling alone (only about 20\% smaller for the SRF case, the laser pulse being 30\,ps rms for the DC gun). The main reason for this is thought to be the $\times3$ difference in laser pulse duration, and as a result the virtual cathode onset condition occurring at a smaller laser spot. It could also be due to the fact that DC fields are more forgiving with respect to time \& energy correlations arising from the virtual cathode condition, whereas the SRF time dependent fields are less amenable to the instability, requiring a larger margin between the theoretical laser spot minimum and the actual spot chosen.

\subsection{Sources of Emittance Growth}

\begin{figure}[t!]
\begin{center}
	\includegraphics[width=0.8\columnwidth]{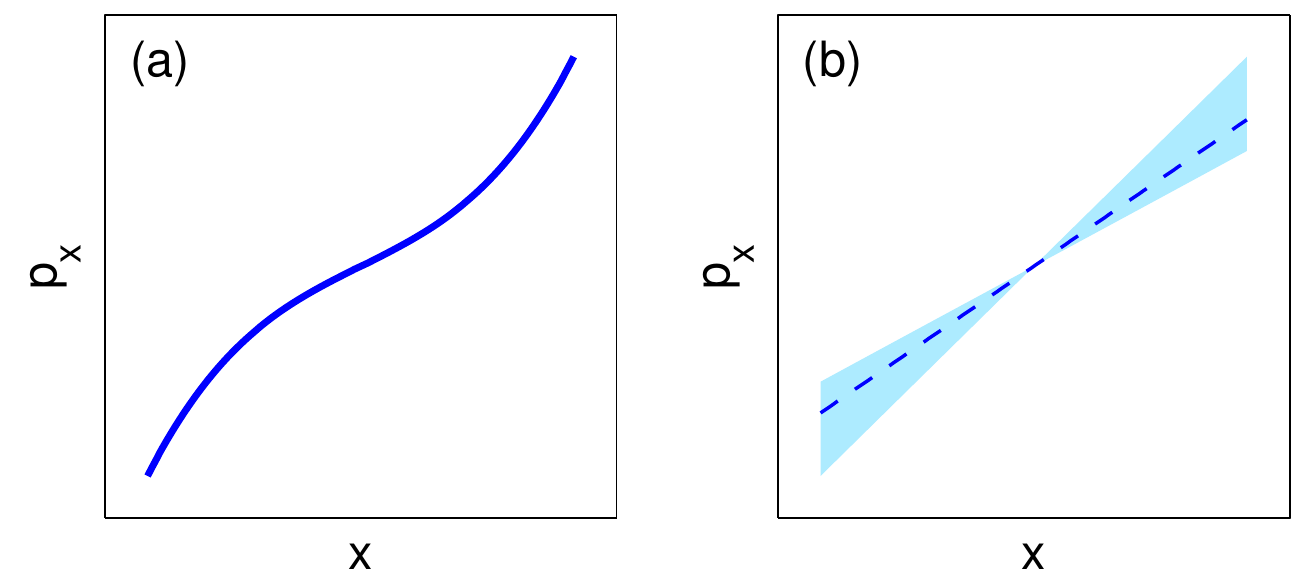}
	\caption{Examples of projected phase space emittance growth from (a) geometric aberrations and (b) projected (slice) phenomena. The dashed line shows an individual slice either in time (for RF-induced emittance) or in energy (for chromatic aberration).\label{fig:cartoon}}
\end{center}
\end{figure}

We differentiate between emittance degrading effects that affect beam slice emittance with those that only increase beam projected emittance, as illustrated in Fig.~\ref{fig:cartoon}. Aberrations in axially symmetric elements arise from $\propto r^3$ dependencies of transverse fields and are commonly referred to as geometric aberrations. The focal length dependence on beam energy (e.g.~in solenoids) or time (in RF cavities) leads to correlated emittance growth which typically does not affect beam slice emittance. This correlated emittance growth can be canceled by other emittance growth mechanisms (RF, space charge, chromatic aberration) that (de)focus beam slices in the opposite direction.

\subsubsection{Solenoid and DC Gun Aberrations}
To evaluate aberrations from magnetic fields in the solenoid, we use the paraxial field expansion to relate the radial component $B_r$ to derivatives of the axial field $B_z$:
\begin{equation}
B_r = -\frac{r}{2}\frac{\partial B_z}{\partial z} + \frac{r^3}{16}\frac{\partial^3 B_z}{\partial z^3} + \mathcal{O}(r^5),
\end{equation}
with an analogous expression existing for the electric field. Given the counteracting effect of space charge, we find that the beam size does not change appreciably inside many of the elements. Thus, we can integrate the transverse momentum imparted by the solenoid using a rigid beam approximation (i.e.~negligible radial coordinate change inside the element) and write
\begin{equation}
r' = \left(\frac{1}{f}\right)_\mathrm{sol} r + \alpha r^3.
\end{equation}
The focal strength has the familiar expression
\begin{equation}
\left(\frac{1}{f}\right)_\mathrm{sol} = \left(\frac{e}{2mc\beta\gamma}\right)^2 \int B_z^2 dz,
\label{eq:sol_focus}
\end{equation}
whereas the aberration (cubic) coefficient $\alpha$ is given by
\begin{equation}
\alpha = \frac{1}{4}\left(\frac{e}{2mc\beta\gamma}\right)^2 \int \left( \frac{\partial B}{\partial z} \right)^2 dz.
\end{equation}

The rms (geometric) emittance growth in Larmor frame can be readily calculated (assuming a zero emittance beam at the entrance):
\begin{equation}
\epsilon_x = \kappa \alpha \sigma_x^4, \label{eq:aber_geom}
\end{equation}
where $\kappa=\sqrt{8}, \sqrt{200/147}$ and $\sqrt{8}/3$ for Gaussian, elliptical, and uniform transverse distributions respectively. The normalized rms emittance due to this geometric aberration for otherwise identical beams scales inversely with the beam momentum, $\epsilon_{nx} \propto 1/(\beta\gamma)$. When the beam is offset from the solenoid magnetic axis by $x_0$, the emittance increase squared is given by
\begin{align}
\epsilon_x^2 & =  4 \alpha^2 \sigma_x^6(5x_0^2+2\sigma_x^2) & \mbox{for Gaussian},\\
\epsilon_x^2 & =  \frac{8}{9} \alpha^2 \sigma_x^6(9x_0^2+\sigma_x^2) &\mbox{for elliptical},\\
\epsilon_x^2 & =  \frac{4}{147} \alpha^2 \sigma_x^6(357x_0^2+50\sigma_x^2) &\mbox{for uniform}.
\end{align}
When $x_0 \ll \sigma_x$ these equations reduce to Eq.~\ref{eq:aber_geom}, whereas for $x_0 \geq \sigma_x$ the emittance increase is given by
\begin{equation}
\epsilon_{x} = \mathcal{K} \alpha \sigma_x^3 x_0.
\end{equation}
Here $\mathcal{K}=\sqrt{20}$, $\sqrt{68/7}$, and $\sqrt{8}$  for Gaussian, elliptical, and uniform transverse distributions respectively.

To estimate geometric abberations from the DC gun, we use numerical integration through the electric field and fit the transverse momentum at the exit of the gun with
$p_x = C x + \alpha_p x^3$, where $x$ is the initial offset on the cathode, $C$ a linear fit coefficient (focusing strength), and $\alpha_p$ the aberration coefficient. The rms normalized emittance in this case is given by analogous to Eq.~\ref{eq:aber_geom} expression:
\begin{equation}
\epsilon_{nx} = \kappa \sigma_x^4 \frac{\alpha_p}{mc}.
\end{equation}
For the DC gun, the coefficient $\alpha_p/mc$ evaluates to $0.029\,\mu\mathrm{m}/\mathrm{mm}^4$ resulting in negligible emittance contribution.

The solenoid has a short focal length, which also leads to strong chromatic effects. Differentiating the focal length, Eq.~\ref{eq:sol_focus}, with respect to beam momentum, we obtain rms normalized emittance increase (in the Larmor frame):
\begin{equation}
\epsilon_{nx} = 2 \left(\frac{1}{f}\right)_\mathrm{sol} \sigma_x \sqrt{\sigma_x^2+x_0^2} \frac{\sigma_p}{mc}.
\end{equation}
For a longitudinally correlated energy spread, chromatic aberrations do not increase slice emittance, as opposed to geometric aberrations, Eq.~\ref{eq:aber_geom}, and therefore can be readily compensated. This effect is significant for both gun types, although it is larger in the DC gun case due to lower energy and larger beam size.

\subsubsection{RF Focusing}
For the RF cavities (SRF gun, RF buncher and energy booster section), the dominant effect tends to be time-dependent RF focusing. The focusing is a function of cavity gradient, phase, and initial beam kinetic energy. No analytical expression exists for RF-focusing in the non-relativistic regime. We numerically obtain the coefficient of the Taylor expansion for the transverse momentum imparted by the cavity, $\partial^2 p_x/ \partial x \partial t$. Overall, the rms normalized emittance contribution is given by
\begin{equation}
\epsilon_{nx}=\frac{1}{mc} \left| \frac{\partial^2 p_x}{\partial x \partial t} \right| \sigma_t \sigma_x \sqrt{\sigma_x^2 + x_0^2},
\end{equation}
$x_0$ being the offset of the beam with respect to the cavity axis, and $\sigma_t$ being the rms bunch duration. This effect is significant for the SRF gun, buncher cavity, and the first 2 SRF cavities.

\subsubsection{Various emittance contributions}
We summarize the various emittance contributions greater than $0.1\,\mu$m for the cases previously depicted in Fig.~\ref{fig:dcenv} and \ref{fig:srfenv} in Table~\ref{tab:aber}.

\begin{table}
\caption{\label{tab:aber}Correlated and uncorrelated emittance contributions (rms normalized in $\mu$m).}
\begin{tabular}{lcc}
Emittance contribution & DC gun & SRF gun\\
\hline
Thermal emittance & $0.17$ & $0.10$ \\
SRF gun (RF induced) & -- & $0.17(0.11)$\\
Buncher (RF induced) & $0.20$ & -- \\
Solenoid1 (geometric) & $0.16[0.13]$ & $<\!0.1$ \\
Solenoid1 (chromatic) & $0.8$ & $0.5$ \\
Solenoid2 (chromatic) & $0.7$ & -- \\
SRF cavity1 & $0.7(0.5)$ & $0.3$\\
\hline
Final transverse emittance &   $0.21$      & $0.15$ \\
\end{tabular}
\end{table}

Geometric aberration are evaluated assuming an elliptical transverse distribution [or uniform in the square brackets]. Where two different values are given in Table~\ref{tab:aber}, the values in parenthesis were obtained with a rigid beam approximation.

It is seen that nontrivial cancelation of correlated emittance growth contributions takes place in both injector types, especially the DC gun based variant. Precision control of the 3D laser shaping and beam optics is required to achieve such a high degree of cancelation.

\section{Conclusions}
We have demonstrated a new optimization method, wherein a genetic algorithm is used to dynamically adjust the gun geometry to achieve the lowest beam emittance.  A comparison of two technologies, DC and SRF guns, for production of high average current low emittance beams has been performed. Undoubtedly both approaches will be pursued by the accelerator community in the coming decade. While each approach has its pros and cons, our optimizations show that either is capable of producing similar quality beams. The analysis performed also emphasizes the importance of low mean transverse energy photocathodes.

\section{Acknowledgements}
This work is supported by NSF DMR-0807731 grant. Andrew Rzeznik implemented the RF cavity autophasing routine in \textsc{gpt}, and Tsukasa Miyajima implemented laser shaping routines. Sergey Belomestnykh and Bruce Dunham are acknowledged for reading and commenting on early versions of the manuscript. CHESS computer group support of the computer cluster is gratefully acknowledged.

\bibliographystyle{elsarticle-num}

\end{document}